\documentclass[usegraphicx]{mn2e}
\begin{document}

\title[Comment]{NGC 2419 does not challenge MOND}

\author[R.H. Sanders] {R.H.~Sanders\\Kapteyn Astronomical Institute,
P.O.~Box 800,  9700 AV Groningen, The Netherlands}

 \date{received: ; accepted: }

\maketitle

\begin{abstract}

I show that, in the context of MOND, non-isothermal models,
approximated by high order polytropic spheres, are 
consistent with the observations of the radial distribution of the
line-of-sight velocity dispersion in the distant globular cluster, 
NGC 2419.  This calls into question the claim by Ibata et al.
that the object constitutes a severe challenge for MOND.
In general, the existence and properties of globular clusters
are more problematic for LCDM than for MOND.  

\end{abstract}

\section{Introduction}

In a recent preprint, Ibata et al. (2011) have claimed that in the
distant globular cluster, NGC 2419, the observed radial profile of
line-of-sight stellar velocity dispersion, combined with the surface
density distribution of starlight, is consistent with Newtonian
gravity in the context of a broad class of globular cluster 
models and inconsistent with
modified Newtonian dynamics (MOND). They go further and
assert that this object is a ``crucible'' for
acceleration-based theories of gravity 
primarily because its large
galacto-centric distance means that the
galactic external field is less than 20\% of the MOND critical
acceleration, $a_0$.  Thus, the complications due to
the MONDian external field effect should be negligible; it is
a relatively isolated system and can be modeled taking only
internal dynamics into account.  

I argue here that the
elevation of this object to the role of crucible is overstated
primarily due to the limitations of the models considered
and that a class of MONDian models which deviate slightly
from an isothermal state are consistent with the
observations.  More generally, I claim that the very existence
and overall properties of globular clusters are more
problematic for LCDM than for MOND.

\section{Background}

The primary models considered by Ibata et al. are Michie models
which are an anisotropic extension of the well known King
models.  These models are based upon the Jeans theorem
which states that, in steady state, the phase space
distribution of stars, $f(\bf{r},\bf{v})$ in any stellar system is
a function of the integrals of motion, in this case energy
and angular momentum.  The Jeans theorem
does not say what the function should be;  this requires
an educated guess.  In spherical systems, a reasonable
supposition, and one justified by observations, is that this
function is an exponential of energy $e^{-E/\sigma^2}$
because this yields a gaussian velocity distribution
with constant velocity dispersion -- an isothermal
sphere.  The problem is that, with Newtonian dynamics,
the isothermal sphere is infinite in extent and mass,
which led Michie (1963, see also King 1966) to impose
of a cutoff radius:  
$f\rightarrow 0$ at some finite
radius corresponding to the tidal limit of the cluster
in the gravitational field of the parent galaxy.  This
is encoded in the distribution function by taking
$f\propto e^{-E/\sigma^2} - 1$ so that the phase space 
density falls to zero where $E=0$,
defined to be the tidal radius.

Long ago, Milgrom (1984) demonstrated that with MOND, 
isothermal spheres have finite mass but infinite
extent;  the density falls asymptotically as $r^{-\alpha}$
where $\alpha \approx 4-5$.  
However, one may show that if the system deviates
from being isothermal (Sanders 2000), even slightly, 
then it is finite
in both mass and extent; the density falls
to zero at a finite radius.  This is true when
deviations from an isothermal state are represented
by high order polytropes,
defined by ${\sigma_r}^2 \propto \rho^{1/n}$ where 
$\sigma_r$ is the radial velocity dispersion, $\rho$ is
the density and $n$ is the polytropic index taken
to be greater than 10 (Newtonian polytropes are finite
in extent only if $n\le 5$).  Such objects naturally contain
a truncation radius which may or may not be identified with a
tidal radius.  High-order polytropes ($12<n<16$)
with a radial velocity anisotropy in the outer regions
can reproduce the general properties of elliptical
galaxies (Sanders 2000) and, in particular, the scaling relations such
as the observed 
fundamental plane and the Faber-Jackson relation; globular
clusters lie on the low mass extension of these relations.  
Therefore, while the polytropic assumption is certainly
an idealization, it would seem 
reasonable to apply such models to globular clusters,
particularly given the existence of an intrinsic truncation
radius.

\section{The non-isothermal model}

The equation solved is the spherically
symmetric Jeans equation
$$ {dP\over {dr}} + {{2\beta P}\over r} = -\rho g \eqno(1)$$
where $P$ is the pressure given by
$$P = \rho{\sigma_r}^2\eqno(2)$$ and
$\beta$ is the anisotropy parameter given by
$$\beta = 1-{\sigma_t}^2/{\sigma_r}^2\eqno(3)$$
with $\sigma_t$ being the velocity dispersion in the tangential
direction.  As is usual I assume a radial dependence
of $\beta$ given by
$$\beta = [1+(r_a/r)^2]^{-1};\eqno(4)$$ that is, for $r<r_a$
the velocity field is isotropic and for $r>r_a$ the stellar
orbits become primarily radial.  Here $g$ is the 
gravitational acceleration given in this case by the MOND formula
$$ g\mu(g/a_0) = g_N\eqno(4)$$
where $g_N$ is the standard Newtonian gravitational acceleration
and $\mu$ is the function interpolating between the Newtonian
regime ($\mu(x)\approx1$ where $x>>1$) and the MOND regime ($\mu(x)\approx x$
where $x<<1$).  In this case, the equations are closed by 
taking the polytropic relation
$${\sigma_r}^2 = {c_0}^2 \Bigl({\rho\over\rho_0}\Bigr)^{1\over{n}}
\eqno(5)$$ where $c_0$ is the central velocity dispersion 
(with MOND this sets the mass of the system), and $\rho_0$ is the
central density. 

For an isothermal sphere or high order polytrope of finite
index, there is a family of solutions characterized by
the form of the central density distribution.  
In the limiting solution of this family the central 
density distribution is power-law in radius -- as a cusp --
but other solutions have a central core of near constant
density:  lower central densities yield larger core radii.

Therefore, the parameters of any such model are
$c_0$, the central velocity dispersion which is essentially
determined by the observed central velocity dispersion; $n$,
the polytropic index which is typically between 10 and 20; $r_a$, the 
anisotropy radius which must be on the order of the
half-light radius to avoid radial orbit instability; $\rho_0$, 
the central density which must be chosen to match the core 
radius.  

\begin{figure}
\resizebox{\hsize}{!}{\includegraphics{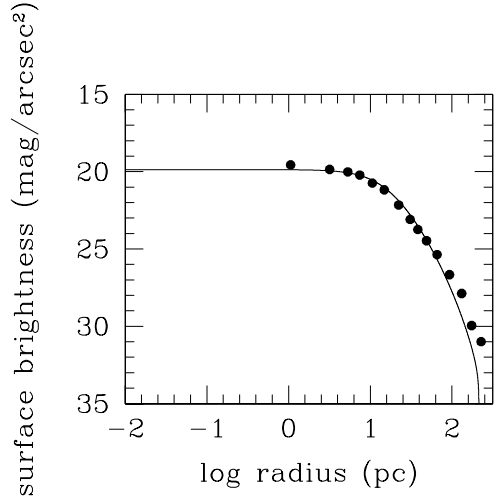}}
\caption{Surface brightness distribution in the
globular cluster NGC 2419 (points) compared to the
MOND non-isothermal, non-isotropic model described in the text.}
\label{}
\end{figure}

\begin{figure}
\resizebox{\hsize}{!}{\includegraphics{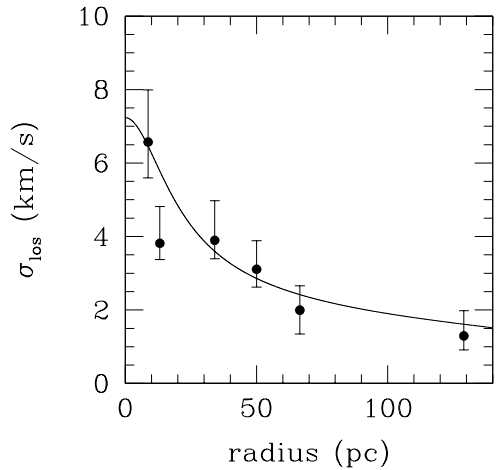}}
\caption{The observed radial dependence of
the line-of-sight velocity dispersion in NGC 2419 (points)
compared to that of the MOND model shown above.
}
\label{}
\end{figure}

The properties of one such model are shown in the two figures
(taking $\mu(x) = x/\sqrt{(1+x^2)}$ and $a_0 = 10^{-8}$
cm/s$^2$).
Fig. 1 is the projected surface brightness distribution 
compared to that observed for NGC 2419.
The second is the radial distribution of line-of-sight velocity
dispersion again compared to the observations.  For this 
model $c_0 = 7.5$ km/s, $n=10$, $r_a = 18$ pc, and $\rho_0=
35$ M$_\odot$ pc$^{-3}$.  This yields a total mass of 
$7.7 \times 10^5$ M$_\odot$ and an effective radius of 18 pc.
The corresponding mass to light ratio is 1.6.
It is evident that this MONDian model provides a reasonable
description of the observations.  Overall, a comparable match
to the observations can be achieved for $7\, {\rm km/s}\le c_0\le 7.5\,
{\rm km/s}$, 
$10\le n \le 12$, and $18\,{\rm pc}\le r_a\le 22\,{\rm pc}$

Ibata et al. have also applied the Jeans equation to consider
a range of non-isothermal, anisotropic models; in these models the density
distribution was frozen to be that of the best-fit Newtonian
model of the cluster because this provides an accurate description
of the surface brightness distribution.  On the basis of the maximum likelihood
ratio test no MOND fit to the data
that is comparable to the best Newtonian model is achieved.  But here
it is evident that the anisotropic polytropes can 
provide a reasonable description of both the surface brightness
distribution and the velocity dispersion radial profile, reasonable
to the eye if not to maximum likelihood.  
This is striking in that the polytropic assumption is a particularly
idealized and rigid method of encoding deviations from an isothermal 
state.  Given this rigidity it is not surprising that the model shown
in Figs. 1 and 2 is not a precise match to the data; in particular,
the predicted surface brightness distribution is an imperfect fit.   
But it is perhaps unwise to rely too heavily on 
formal statistical tests which assume random errors in data
that may be plagued by systematic effects, observational and/or
physical (a common problem in the interpretation astronomical data).

\subsection{Discussion}

Observationally, there are essentially
two classes of halo objects:  globular clusters and
dwarf spheroidal galaxies; these overlap in mass but
not in surface brightness or in age and uniformity
of the stellar populations.  The globular clusters are
generally comprised of ancient, though not necessarily
coeval, stellar populations
and they are numerous (several hundred observed in and
inferred in the Galaxy).  They are high surface brightness objects and
show no dynamical evidence for dark matter within the
visible object (dynamical mass-to-light ratios 
are typical of the observed
stellar populations).  The dwarf spheroidal galaxies
have low surface brightness and a large conventional
dynamical M/L (M/L exceeding 100 in some cases).
They are few in number (about 20 directly observed in the
the Galaxy) and contain generally younger stellar populations
covering a range of ages. 

In the context the LCDM paradigm,
the explanation of the general properties of these halo objects,
specifically the presence or absence of ``dark matter",
resides in murky creation mythology.  LCDM simulations predict
that galaxies, assembled over cosmic time via mergers of
smaller halos, should contain a large number of dark matter
sub-halos (in the Galaxy more than 200 with velocity dispersion 
greater than a few km/s);  this substructure is intrinsic
to the theory and a fundamental constituent of galaxy scale
halos.  It might seem more natural to identify
these dark matter sub-halos with globular clusters,
the more numerous and primordial objects in the Galaxy.
However, these are baryon rather than dark matter dominated,
so the identification is made with the dwarf spheroidals. Their
embarrassing scarcity is then due the fact that most
of the predicted dark matter sub-halos have remained dark
because they never
captured sufficient baryons to initiate star formation or the
captured baryons have been blown away by early stellar processes. 
Then a separate formation scenario must be invoked
for globular clusters; e.g., globular clusters
are formed in primordial disk-bound supermassive molecular
clouds with high baryon to dark matter ratio and later attain
a more spheroidal shape due to subsequent mergers (Kravtsov \&
Gneden 2005).  
These scenarios, while imaginative, are, to say the least, 
difficult to falsify.

In the context of MOND there is
no need to speculate about formation processes in order
to account for the perceived dark matter content in
these two classes of objects.
MOND predicts that high surface brightness systems
(systems with high internal acceleration) should exhibit
no evidence for a mass discrepancy within the visible
object (conventionally, no dark matter).
Conversely, MOND predicts that low-surface-brightness systems,
such as the dwarf spheroidal satellites of the
Galaxy, should exhibit a large discrepancy.  
These observed properties of globular clusters and
dwarf spheroidals find natural
explanation in the context of MOND based on existent 
physical law, not on formation scenarios.  That is not to say differing
formation histories are unimportant in defining
the overall observed properties of these two distinct
classes of halo objects (such as the stellar populations).
With MOND, globular clusters could
well be among the first objects formed, prior to or simultaneous
with galaxies,
as suggested by the old stellar populations; whereas
the dwarf spheroidals may have formed subsequently as
tidal objects.  But the magnitude of
the apparent "dark matter content" is directly related
to the internal acceleration or observed surface density and not
to different formation histories.

It is of interest that with MOND, non-isothermal systems,
such as the high-order polytropes shown here, have a cutoff radius
(an edge) which is unrelated to the tidal radius.  Given the
baryonic mass of NGC 2419, the tidal radius should be in excess
of 1 kpc, and yet, the observed truncation radius is on the order of 200 pc.
In general the cutoff radii of dwarf spheroidals, which have
comparable baryonic masses, are larger than those of the
globular clusters (Zhao 2005a,b).  Perhaps it is the case
that the globular clusters do not fill their Roche lobe -- that
the density cutoff is due to non-isothermal state.  On the other
hand, the dwarf spheroidals may well extend to their tidal
radii because of the different formation history.

With respect to the specific example of NGC 2419 it
has been claimed that simultaneously matching the 
radial distribution of starlight and line-of-sight velocity
dispersion 
is not possible in the context of MOND.  This claim
is made in the context of a class of  
isothermal models in which the phase space distribution of
stars as a function of the integrals of motion is chosen
to be of a quite specific form (the Michie model).  
This class may be appropriate for Newtonian isothermal spheres with
a constructed radial cutoff (identified with the tidal radius) but it is
not clearly applicable to MONDian objects which
are intrinsically finite.  I have presented a counter-example which
demonstrates no such inconsistency with the observations:  
a non-isothermal models, approximated by high order
polytropes.  I attach no particular significance to the
polytropic relation between velocity dispersion and density;
it is an idealized assumption.  But it does demonstrate that, given the
uncertainties of anisotropy or isothermality, it is perhaps
rash to claim that this one particular object
is problematic for MOND.  Each single well-measured
rotation curve for a nearby disk galaxy -- missing 
these ambiguities -- is far more of a crucible for
gravity theories.

I am grateful to Moti Milgrom for helpful comments on the
draft of this paper.  I also thank Roderigo Ibata for
pointing out an inconsistency in earlier results that
led to a correction of my Jeans equation integrator.
And I thank Scott Trager for a useful discussion on
the properties of globular clusters.

\end{document}